# A SURVEY OF PRESSURE VESSEL CODE COMPLIANCE FOR SUPERCONDUCTING RF CRYOMODULES

Thomas Peterson[1], Hitoshi Hayano[2], Kay Jensch[3], Eiji Kako[2], Arkadiy Klebaner[1], John Mammosser[4], Axel Matheisen[3], Hirotaka Nakai[2], Tom Nicol[1], Jay Theilacker[1], Akira Yamamoto[2]

1. Fermi National Accelerator Laboratory, Batavia, IL (USA)
2. KEK High Energy Accelerator Research Organization, Tsukuba, Ibaraki, 305-0801, Japan
3. Deutsches Elektronen Synchrotron, DESY, Hamburg, 22607, Germany
4. Thomas Jefferson National Accelerator Facility, Newport News, VA (USA)

## ABSTRACT

Superconducting radio frequency (SRF) cavities made from niobium and cooled with liquid helium are becoming key components of many particle accelerators. The helium vessels surrounding the RF cavities, portions of the niobium cavities themselves, and also possibly the vacuum vessels containing these assemblies, generally fall under the scope of local and national pressure vessel codes. In the U.S., Department of Energy rules require national laboratories to follow national consensus pressure vessel standards or to show "a level of safety greater than or equal to" that of the applicable standard. Thus, while used for its superconducting properties, niobium ends up being treated as a low-temperature pressure vessel material. Niobium material is not a code listed material and therefore requires the designer to understand the mechanical properties for material used in each pressure vessel fabrication; compliance with pressure vessel codes therefore becomes a problem. This report summarizes the approaches that various institutions have taken in order to bring superconducting RF cryomodules into compliance with pressure vessel codes.

## INTRODUCTION

Superconducting radio frequency (SRF) cavities made from niobium and cooled with liquid helium are becoming key components of many particle accelerators. These SRF cavities are typically cooled to low temperatures by direct contact with a liquid helium bath, resulting in at least part of the helium container being made from pure niobium and/or niobium-titanium. In the U.S., Europe, and Japan, these helium containers and part or all of the RF cavity fall under the scope of the local and national pressure vessel rules. Thus, while used for its superconducting properties, niobium must be treated as a material for pressure vessels. Problems with the certification of pressure vessels constructed partially or completely of niobium arise due to the fact that niobium and titanium are not listed as an acceptable vessel materials in pressure vessel codes. Within the ASME code, in particular, pure niobium is not approved for use in Division 1 or Division 2 vessels [1], and there are no mechanical properties available from code sources for either niobium or titanium at liquid helium temperature. Thus, showing a level of safety greater than or equal to that of the applicable standard, as is typically required when one cannot entirely meet code requirements, involves not only following code design, inspection, and documentation rules as much as possible, but also establishing a safely conservative set of niobium mechanical properties for the vessel and doing detailed analyses of niobium stresses.

We examine here how various organizations around the world plan to meet requirements as closely as possible for compliance with pressure vessel rules in the design of SRF helium vessels including niobium RF cavities.

**APPLICATION OF THE PRESSURE VESSEL CODES**

Cavity design that satisfies level of safety equivalent to that of a consensus pressure vessel code is affected by use of the non-code material (niobium), complex forming and joining processes, a shape that is determined entirely by cavity RF performance, a thickness driven by the cost and availability of niobium sheet, and a possibly complex series of chemical and thermal treatments.

FIGURE 1 illustrates one configuration of a niobium SRF cavity within a titanium helium vessel. Liquid helium surrounding the niobium SRF cavity exerts an internal pressure on the surrounding helium jacket and head components, and an external pressure on the niobium cavity itself. This pressure may occur both at ambient temperatures with gaseous helium, for example during purging and cool-down and at cryogenic temperatures. The end parts and joints of the surrounding helium vessel experience tensile stress while at least portions of the niobium cavity are susceptible to buckling failure. Welding techniques to connect niobium parts of the RF cavity were developed to provide optimum surface conditions at the welds of the resonators exposed to radio frequencies. The welding technique applied for connecting the helium vessel and Nb resonator is optimized for minimum influence on the resonator frequency. The helium vessel must be hermetically closed, leak tight. These welds are not designed and optimized with respect to PED weld codes.

In applying any of the consensus pressure vessel or piping codes, key factors to demonstrate the required level of design safety are the establishment of a maximum allowable stress, and (for external pressure design) an accurate approximation of the true stress strain curve [2]. Then, due to the complex geometry of SRF cavities and the surrounding helium vessels, detailed analysis is required to determine stresses as a function of pressure and to verify a maximum allowable working pressure (MAWP).

Weld inspections by visual or radiography methods may not always be possible in accordance with pressure vessel code requirements due to the assembly procedures, use of electron beam welding, cleanliness requirements limiting access, and shadowing of welds inside the vessel.

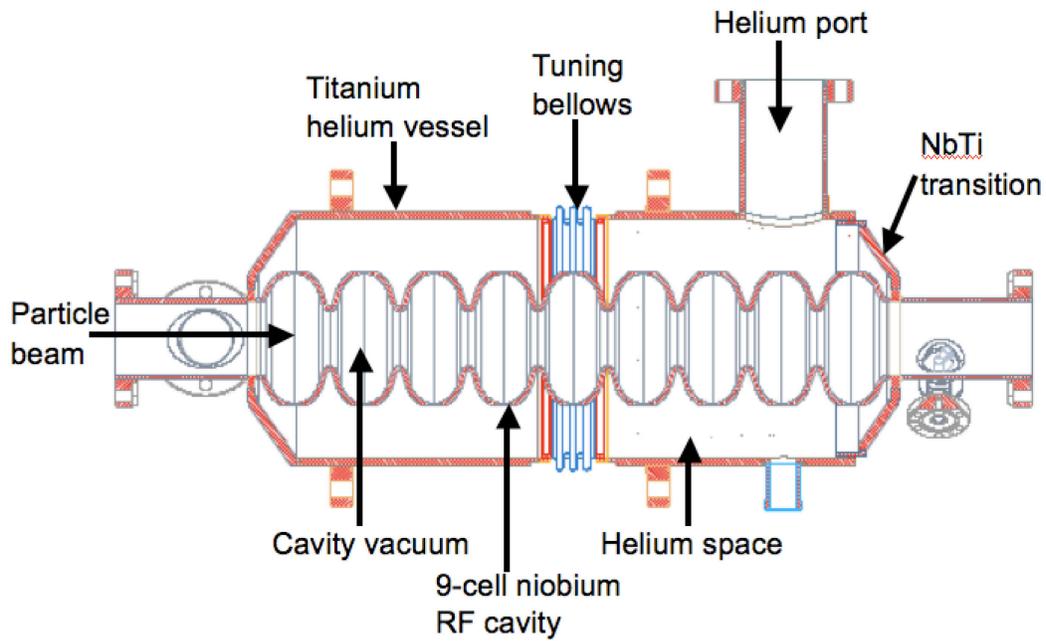

**FIGURE 1.** Illustration of a niobium, multi-cell, elliptical-shape SRF cavity within a helium vessel.

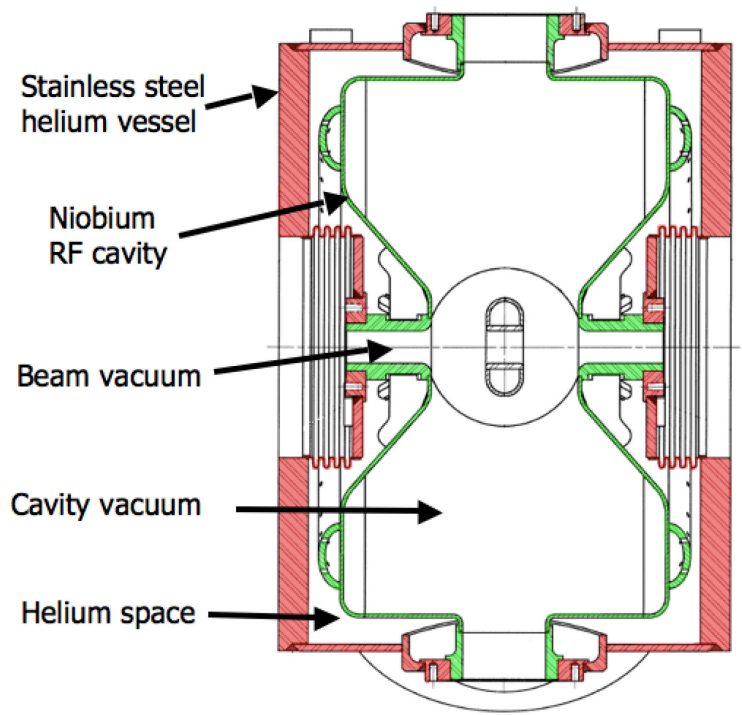

**FIGURE 2.** Illustration of a single-spoke niobium SRF cavity within a helium vessel.

## A SURVEY OF PRESSURE VESSEL COMPLIANCE APPROACHES TAKEN BY ORGANIZATIONS AROUND THE WORLD

Accelerators and R&D facilities around the world have recently been addressing the issue of bringing SRF helium vessels and cavities into compliance with pressure rules.

Various approaches have been taken to show that these vessels are safe not only for personnel but that equipment failure is very unlikely. The following is an overview of some of these approaches to this problem.

**Fermilab (USA)**

In to assure the safety of SRF vessels and to comply with U.S. Department of Energy regulations [3], Fermilab formed a committee to develop a SRF design guideline [4, 5, 6] that provides equivalent protection and a level of safety afforded through ASME. The Fermilab SRF design guideline addresses the unique SRF pressure safety challenges and the use of special materials. The below referenced policy and design guideline will also document compliance with applicable ASME standards as well as the requirements of 10CFR 851, "Worker Safety and Health Program". The two governing Fermilab documents developed to supplement the ASME/ANSI standards are:

Policy: "Dressed Niobium SRF Cavity Pressure Safety", FESHM chapter 5031.6. [5]
Design guideline: "Guidelines for the Design, Fabrication, Testing and Installation of SRF Nb Cavities", Technical Division Technical Note TD-09-005. [6]

To meet the intent of applicable ASME standards and 10CFR 851, the above referenced documents include means to assure the following:
- Design drawings, sketches, and calculations are reviewed and approved by qualified independent design professionals.
- Only qualified personnel must be used to perform examinations and inspections of materials, in-process fabrications, non-destructive tests, and acceptance tests.
- Documentation, traceability, and accountability is maintained for each pressure vessel and system, including descriptions of design, pressure conditions, testing, inspection, operation, repair, and maintenance.

The more detailed "Guidelines for the Design, Fabrication, Testing and Installation of SRF Nb Cavities" document [6] outlines the following requirements for the documentation and certification of the dressed cavity:
- Cavity description
- Material data
    - Materials and properties used in construction
    - Material certifications
    - Serial numbers of cells (traceability)
- Design calculations
- Fabrication information
    - Welding / brazing details and specifications
    - Welder's qualification
    - Processing history
- Cavity wall thickness
    - Internal pressure
    - External pressure
- Examination reports
- Pressure test reports
- Relief system verification
- Operating procedures

In addition, a new SRF review panel has been formed by Fermilab's Cryogenic Safety and Mechanical Safety Subcommittees to provide an independent review of new SRF designs for compliance to the policy and design guideline.

The Fermilab standard only applies to "dressed cavities", defined as "An integrated assembly wherein a niobium cavity has been permanently joined to a cryogenic containment vessel, such that niobium is part of the pressure boundary and the cavity is surrounded by cryogenic liquid during operation." The tests of bare niobium cavities, not yet integrated with a helium vessel but instead in a test dewar, fall under the scope of existing standards which pertain to the test apparatus.

**Brookhaven (USA)**

Engineers at Brookhaven, Advanced Energy Systems (AES) and Stony Brook University have analyzed cavity vessel stresses in accordance with ASME code rules in order to satisfy code requirements. "Through the use of [ASME Boiler and Pressure Vessel Code] Div. 2 requirements and sound engineering judgment an equivalent level of safety to the ASME Boiler and Pressure Vessel Code can be achieved." [8] They have applied this approach also to the Cornell SRF cavity design "CESR-B", which is now used in several particle accelerator facilities around the world. [9]

**Spallation Neutron Source (SNS) at Oak Ridge (USA)**

The cryostats containing SRF cavities for the Power Upgrade Project (PUP) at SNS at Oak Ridge National Lab in the US will comply with pressure vessel rules by treating the vacuum vessel as the pressure boundary. Since the vacuum vessel may primarily be made of stainless steel and/or other conventional vessel materials, it may be code-stamped.

Advantages of moving the pressure boundary from the helium circuit to the vacuum vessel include
- Stainless steel is a code listed material
- Properties show very high toughness at low temperatures
- Components could easily be fabricated by commercial code shop
- No pressure testing would be required of the helium circuit
- Temperature of the vacuum vessel when a failure occurs will never reach 2K

Thus, by treating the vacuum vessel as the pressure boundary, SNS engineers can avoid the issues associated with the use of niobium as a pressure vessel material. If a niobium cavity fails, the code-stamped vacuum vessel is certified to contain the ruptured vessel and any contents which may have spilled. [10]

**DESY (Germany)**

Deutsches Elektronen Synchrotron (DESY), in Hamburg, Germany, like the U.S. laboratories, was not a licensed manufacturer of pressure vessels when the XFEL project started, but now they are. Except for some R&D vessels, their SRF dressed cavity vessels are produced in industry. DESY has developed pressure vessel criteria jointly with the safety authorities and the cavity vendors. DESY has been able to satisfy European Pressure Directive (PED) [7] regulations in part by virtue of their experience with these SRF vessels.
- Operational experiences over 15 years
- About 160 cavities built

- Long term experience with cold operation of these niobium RF cavities
- Experiences with pressure tests and "crash test" at 2 Kelvin on Cavity C 26 and Module 3
- Traceability of drawings, fabrication, materials and preparation

According to PED criteria, individual cryomodules would be only category II pressure vessels due to their relatively small helium volume and pressure, but connected as they will be in series, they become category IV.

For the XFEL dressed cavities, DESY is defined as the manufacturer. DESY chose the module B and module F of the applicable conformity evaluation criteria in the PED for the production of the cavities dressed with titanium helium tanks. DESY must do finite element calculations basing on the design and fabrication drawing as well as on worst case scenarios of fabrication errors that may occur in production, A "test piece" (pre-production welding test according to ISO 15613) had to be fabricated for destructive tests to qualify the welds and welding technique applied. DESY must set up a PMA (Particular Material Appraisal per PED (97/23/EC) annex I, Sec 4.2) for materials and follow this PMA strictly. To complete the module B (PED) directives DESY will manufacture some number of cavities with pressure vessel code authorities present and do nondestructive testing on them. During the production DESY must strictly follow these PMAs and the welding technique as defined and qualified in module B. No changes are allowed and full traceability has to be given. In accordance with Module F requirements, here a statistical number of cavities are selected from the production where the production records are intensively checked and all weld areas are inspected and proven for their compliance with the data fixed in the module B production qualification. A pressure test at 6.7 bar abs on the completed cavity is performed for each cavity built and has to be performed with a representative of a notified body as witness.

**KEK (Japan)**

Like in the U.S. and Germany, KEK laboratory in Japan has had to satisfy pressure vessel rules in fabricating their SRF dressed cavities. Gaining approval from the pressure vessel authorities was eased by means of the definition of the pressure vessel to exclude the 2-phase pipe. (In FIGURE 1, this pipe would be connected to the nozzle from the helium vessel.) The advantage comes with the category of the pressure vessel having pressure times volume (PV) at PV < 0.004. Process and inspection in accordance with the simplified version of the high pressure code requires material mechanical evaluation prior to the production process. A high pressure code test (pressure and leak test) is only required for the complete assembly.

**SUMMARY**

Fermilab has developed standards for dressed SRF cavities, which when followed, provide a level of safety equivalent to that provided by the ASME Boiler and Pressure Vessel Code. Brookhaven has taken an analytical approach to show an equivalent level of safety.

DESY has gained approval of pressure vessel authorities for their SRF pressure vessels partly by means of their considerable experience and previously safe record of operation of these kinds of vessels and by application of the European Pressure Directive (PED).

KEK must follow the Japanese vessel code but has been able to take advantage of the small helium volume of the 1.3 GHz dressed cavities in order to use a simplified version of the code.

SNS at Oak Ridge has taken a different, unique approach, of code-stamping the vacuum vessel so as to avoid the issue of treating niobium as a pressure vessel material.

## CONCLUSION

In Japan, Germany, and the U.S., institutions building superconducting RF cavities integrated in helium vessels or procuring them from vendors have had to deal with pressure vessel requirements being applied to SRF vessels, including the niobium and niobium-titanium components of the vessels. While niobium is not an approved pressure vessel material, data from tests of material samples provide information to set allowable stresses. By means of procedures which include adherence to code welding procedures, maintaining material and fabrication records, and detailed analyses of peak stresses in the vessels, or treatment of the vacuum vessel as the pressure boundary, research laboratories around the world have found methods to demonstrate and document a level of safety equivalent to the applicable pressure vessel codes.

## ACKNOWLEDGMENTS

Work at Fermilab supported by the U.S. Department of Energy under contract No. DE-AC02-07CH11359.